\documentclass[letter]{aa} 

\usepackage[varg]{txfonts}
\usepackage{multirow}
\usepackage{hyperref}
\usepackage{inputenc}
\usepackage{caption}

\hypersetup{
    colorlinks=true,
    linkcolor=magenta,
    filecolor=magenta,      
    urlcolor=violet,
    citecolor=blue
}

\begin{document}

\setlength{\abovedisplayskip}{3pt}
\setlength{\belowdisplayskip}{3pt}

\title{Parsec-scale properties of the radio brightest jetted AGN at $z > 6$}

\titlerunning{VLBI properties of the radio brightest jetted AGN at $z > 6$}

\author{C.~Spingola\inst{1,2}\thanks{E-mail:\href{mailto:spingola@ira.inaf.it}{spingola@ira.inaf.it}} 
\and D.~Dallacasa\inst{1,2}
\and S.~Belladitta\inst{3,4}
\and A.~Caccianiga\inst{3}
\and M.~Giroletti\inst{2}
\and A.~Moretti\inst{3}
\and M.~Orienti\inst{2}} 

\institute{Dipartimento di Fisica e Astronomia, Universit\`{a} degli Studi di Bologna, Via Gobetti 93/2, I$-$40129 Bologna, Italy
 \and INAF $-$ Istituto di Radioastronomia, Via Gobetti 101, I$-$40129, Bologna, Italy
\and INAF $-$ Osservatorio Astronomico di Brera, via Brera, 28, 20121 Milano, Italy
  \and DiSAT, Università degli Studi dell’Insubria, Via Valleggio 11, 22100 Como, Italy} 
 
\date{Received 17 September 2020  / Accepted 20 October 2020}

\abstract{
We present Director's Discretionary Time multi-frequency observations obtained with the Jansky Very Large Array (VLA) and the Very Long Baseline Array (VLBA) of the blazar PSO J030947.49+271757.31 (hereafter PSO~J0309+27) at $z = 6.10\pm0.03$. The milliarcsecond angular resolution of our VLBA observations at 1.5, 5 and 8.4 GHz unveils a bright one-sided jet extended for $\sim500$ parsecs in projection. This high-z radio-loud AGN is resolved into multiple compact sub-components, embedded in a more diffuse and faint radio emission, which enshrouds them in a continuous jet structure. We derive limits on some physical parameters directly from the observable quantities, such as viewing angle, Lorentz and Doppler factors. 
If PSO~J0309+27 is a genuine blazar, as suggested by its X-ray properties, then we find that its bulk Lorentz factor must be relatively low (less than 5).  Such value would be in favour of a scenario currently proposed to reconcile the paucity of high-z blazars with respect to current predictions. 
Nevertheless, we cannot exclude that PSO~J0309+27 is seen under a larger viewing angle, which would imply that the X-ray emission must be enhanced, for example, by inverse Compton with the Cosmic Microwave Background. More stringent constraints on the bulk Lorentz factor in PSO~J0309+27 and the other high-z blazars are necessary to test whether their properties are intrinsically different with respect to the low-z blazar population.

}

\maketitle

\section{Introduction}
\label{sec:introduction}

Little is observationally known above redshift $z=6$, when the Universe was young and the first sources (including active galactic nuclei, AGN) ionised their surrounding gas in the period called \textsl{cosmic reionization} \citep[e.g.,][]{Zaroubi2013}. The AGN detected at these cosmological distances have already masses higher than $10^{8-9}$ M$_\odot$ \citep[e.g.,][]{Vito2019} which are indicative of a fast and efficient growth that challenges supermassive black holes (SMBH) standard formation models \citep[e.g.,][]{Volonteri2012, Wu2015}.  Among the high-z AGN those that are also radio-loud are about 10 \% of the entire AGN population \citep{Banados2015, Padovani2017}, and provide a unique opportunity to study the role of jets in the accretion of SMBH \citep[e.g.,][]{Volonteri2015}, their feedback on the host galaxy \citep[e.g.,][]{Fabian2012}, the cosmic evolution of the AGN radio luminosity function \citep{Padovani2015} out to the largest distances and can be used as cosmological probes \citep[e.g.,][]{Gurvits1999}.

The radio-loud AGN called \textsl{blazars} have their relativistic jets oriented along the line of sight \citep{Urry1995}. Since their non-thermal radiation is relativistically amplified, and not obscured along the jet direction, blazars are very bright and visible up to high redshifts, allowing the study of the radio-loud AGN population across cosmic time \citep[e.g.,][]{Ajello2009,Caccianiga2019}. On milliarcsecond-scales (probed by very long baseline interferometry, VLBI) blazars usually show an unresolved flat-spectrum component ("core") and a one-sided relatively compact steep-spectrum jet. Moreover, with VLBI it is possible to measure some physical properties (e.g., the brightness temperature, the bulk Lorentz factor, see Appendix \ref{sec:appendix_formulae}) of this class of AGN  \citep[e.g.,][]{Lobanov2010, Lister2013}. 
Before \citet{Belladitta2020}, only 7 blazars were known at $z>5$, the most distant being at $z=5.47$ \citep{Romani2004}. This is essentially due to the rarity of this type of  radio-loud AGN: we estimated $\sim2\times10^{-4}$ blazars every square degree at $z>5.5$ and with flux densities larger than a few mJy, corresponding to about 2--3 blazars between $z=5.5$ and $z=6.5$ detectable in currently available wide field surveys (\citealt{Caccianiga2019}, but see also \citealt{Padovani2017}).

There is a mismatch of at least an order of magnitude between the number of observed blazars at high redshifts and those expected from cosmological models \citep{Haiman2004, Volonteri2011}. Possible solutions to such a large discrepancy include an evolution of the bulk Lorentz factor $\Gamma$ (decreasing for increasing redshift), a radial velocity structure of the jet (namely only part of the jet is actually highly beamed), a compact and self-absorbed structure (that prevents the detection at GHz frequencies), possible observational biases related to strong optical absorption and compact self-absorbed radio emission \citep{Volonteri2011}. In order to clearly assess the origin of this disagreement there is the urgency of detecting and confirming the presence of blazars at large redshifts.

By combining the NRAO VLA Sky Survey catalog (NVSS, \citealt{Condon1998}, in the radio), the Panoramic Survey Telescope and Rapid Response System (Pan-STARRS PS1, \citealt{Chambers2016}, in the optical) and the AllWISE Source Catalog (WISE, \citealt{Wright2010}; NEOWISE, \citealt{Mainzer2011}, in the mid-infrared) and using the \textsl{dropout} technique we discovered PSO~J0309+27, which we spectroscopically confirmed to be at a redshift $z=6.10\pm0.03$ \citep{Belladitta2020}. 
It is the  brightest AGN in the radio band  discovered to date above redshift 6 (23 mJy as measured with the NVSS). Through a dedicated \textsl{Swift}-XRT observation we measured an X--ray flux of $\sim3.4\times10^{-14}$ erg s$^{-1}$ cm$^{-2}$ in the [0.5-10] keV energy band, which makes PSO~J0309+27 also the second brightest AGN in the X-ray band discovered at $z>6$ (with the brightest being CFHQS J142952+544717 from recent \textsl{SRG/e-ROSITA} observations, \citealt{Medvedev2020}). Besides the NVSS observation, PSO~J0309+27 is also detected in the TIFR GMRT Sky Survey (TGSS, $64.2\pm6.4$~mJy, \citealt{Intema2017}) at 147 MHz and at 3~GHz by the VLA Sky Survey (VLASS, $12.0\pm1.2$~mJy, \citealt{Lacy2019}). On the basis of the multi-wavelength observed properties, in particular those at X-ray energies, \citet{Belladitta2020} classified PSO~J0309+27 as a blazar.

In this Letter, we present a follow-up of PSO~J0309+27
with the VLA and the VLBA, which aims at constraining the radio spectral properties of the target and simultaneously investigating the morphology and physical conditions from arcsec to milliarcsecond (mas) scales, thus unveiling the parsec-scale emission of this high-z AGN.
Throughout this paper, we assume $H_0$ = 67.8 km s$^{-1}$ Mpc$^{-1}$, $\Omega_M$ = 0.31, and $\Omega_{\Lambda}$= 0.69 \citep{Planck2016}. The synchrotron spectral index is defined as $S_{\nu} \propto \nu^{\alpha}$. At $z=6.10\pm0.03$ the luminosity distance is $60132 \pm 345$ Mpc, which gives a scale of $5.77 \pm 0.02$~pc mas$^{-1}$.

\section{Observations and data reduction}
\label{sec:observations}
\subsection{VLBA}

We observed PSO~J0309+27 with the VLBA on April 6th, 2020 at central observing frequencies of 1.5, 5 and 8.4 GHz, each with bandwidth of 512 MHz, and with a 2 Gbps data recording rate  under the Director’s Discretionary Time (DDT) project BS294 (PI: Spingola). The observation strategy was that of standard phase referencing, which includes scans on the targets of $\sim$3.5 minutes each interleaved by shorter scans on the phase calibrator (J0316+2733). In the observing run we also included a few scans on  DA193 for the bandpass calibration. The total observing time was of about 2~h for each band. We systematically cycled through the observing frequencies in the 6~h observation in order to obtain the best possible \textsl{uv}-coverage at each band. The correlation was performed using the VLBA DiFX correlator in Socorro \citep{Deller2011}.

\begin{figure*}
    \centering
    \includegraphics[width = 1.01\textwidth]{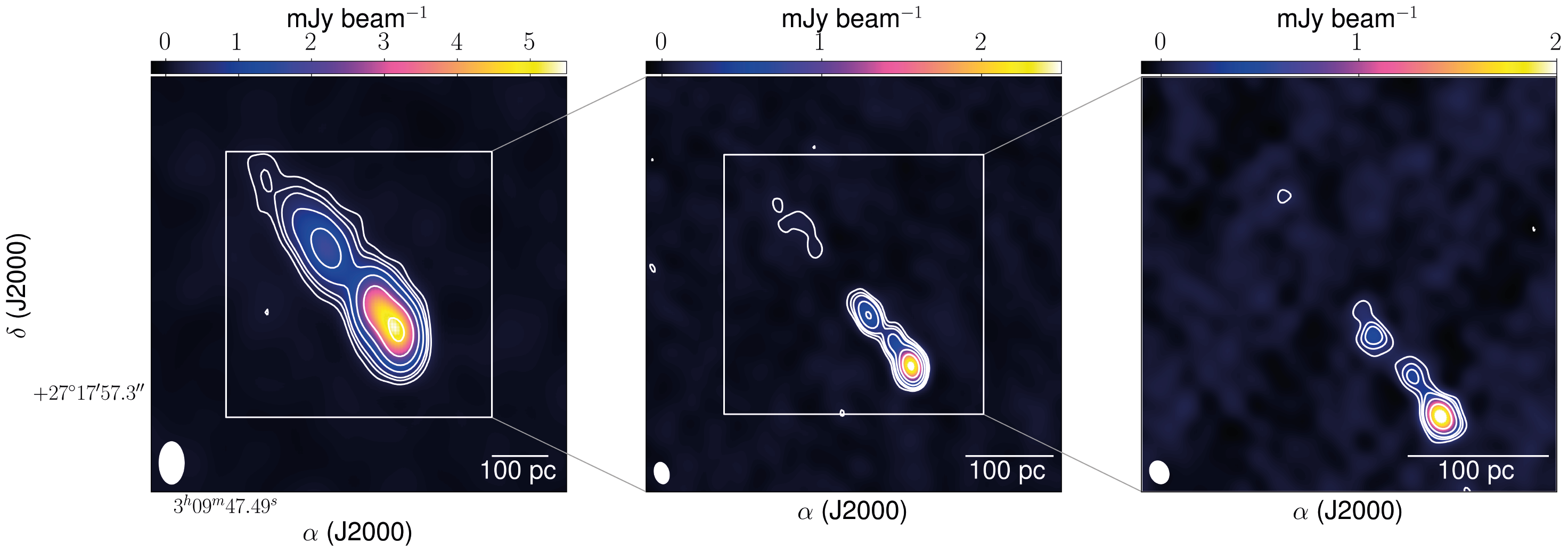}
    \caption{Self-calibrated images of PSO~J0309+27 at 1.5 GHz (left), 5 GHz (center) and 8.4 GHz (right). The contours are drawn at ($-$3, 3, 6, 9, 18, 36, 72, 144) $\times$ the off-source noise of each map, which are given in Table \ref{tab:images_properties}. The white colour bar indicates 100~pc in projection at $z = 6.1$. The restoring Gaussian beam is shown in white in the bottom left corner of each image; north is up, east is left.
    }
    \label{fig:VLBA_images}
\end{figure*}

The data were processed with the Astronomical Image Processing System ({\sc aips}, \citealt{Greisen2003}) package following the standard VLBA calibration procedure for phase-referenced observations, which we briefly summarise.  First, we inspected the visibilities to search for potential bad data (e.g., RFIs, off-source time stamps etc.). Then we corrected for ionospheric dispersive delay, voltage offsets in the samplers, instrumental delay and parallactic angle variation. We then applied a bandpass calibration and performed the \textsl{a priori} amplitude calibration, which uses gain curves and system temperatures. Finally, a global fringe fitting run was performed to correct for the residual fringe delays and rates of the calibrator visibilities. The solutions were then interpolated to the target visibilities. We performed several iterations of phase and amplitude self-calibration, as the solutions had sufficient signal-to-noise ratio, until the rms noise limit was reached.
The final self-calibrated images at the three observing frequencies are shown in Fig.~\ref{fig:VLBA_images}. We adopted a natural weighting scheme to better recover the most diffuse structure of this source at all frequencies. The properties of the images, such as off-source rms noise level, total flux density, peak surface brightness and restoring beam are listed in Table~\ref{tab:images_properties}.

\begin{figure*}
    \centering
        \includegraphics[width = 0.49\textwidth]{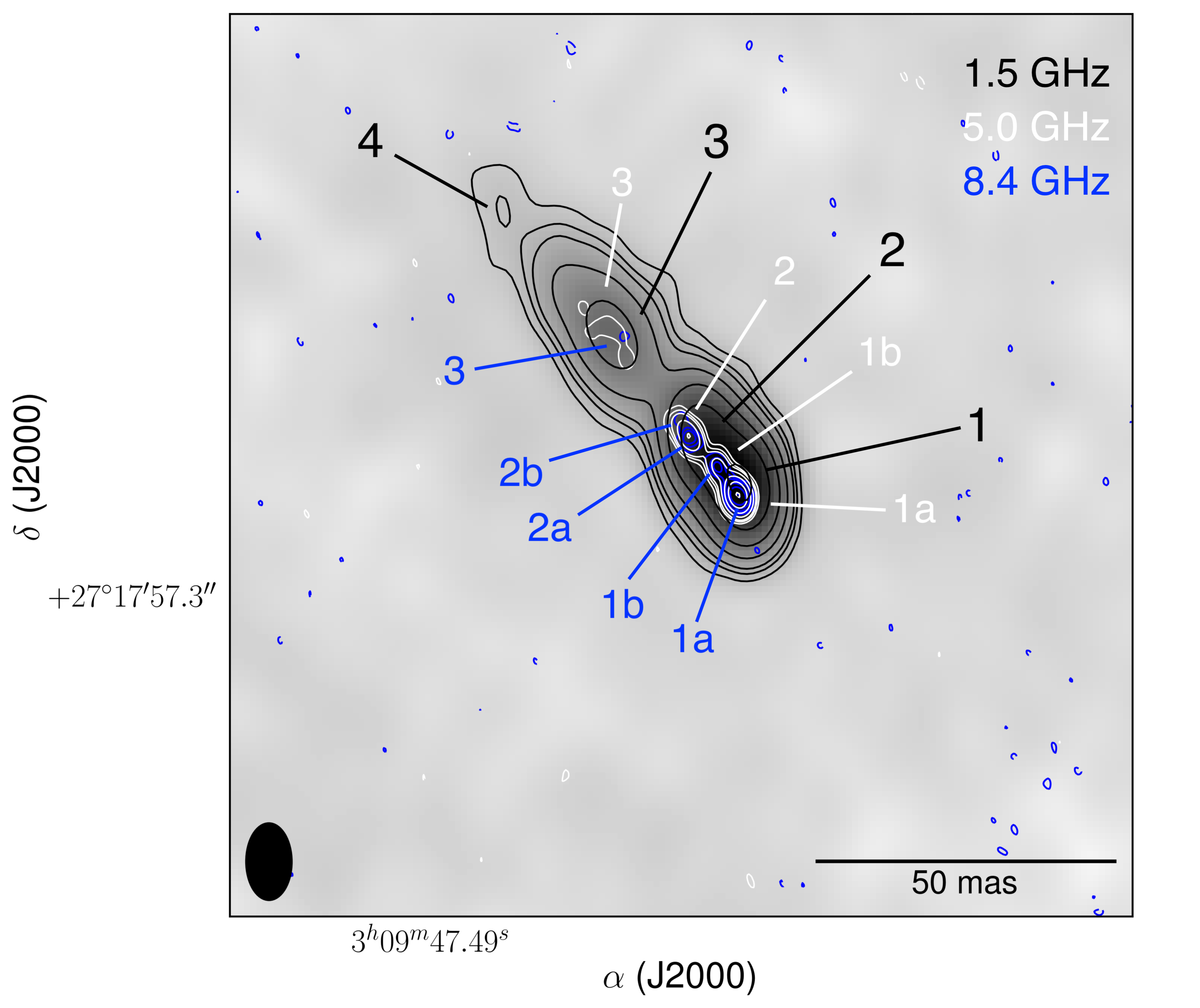}
    \includegraphics[width = 0.46\textwidth]{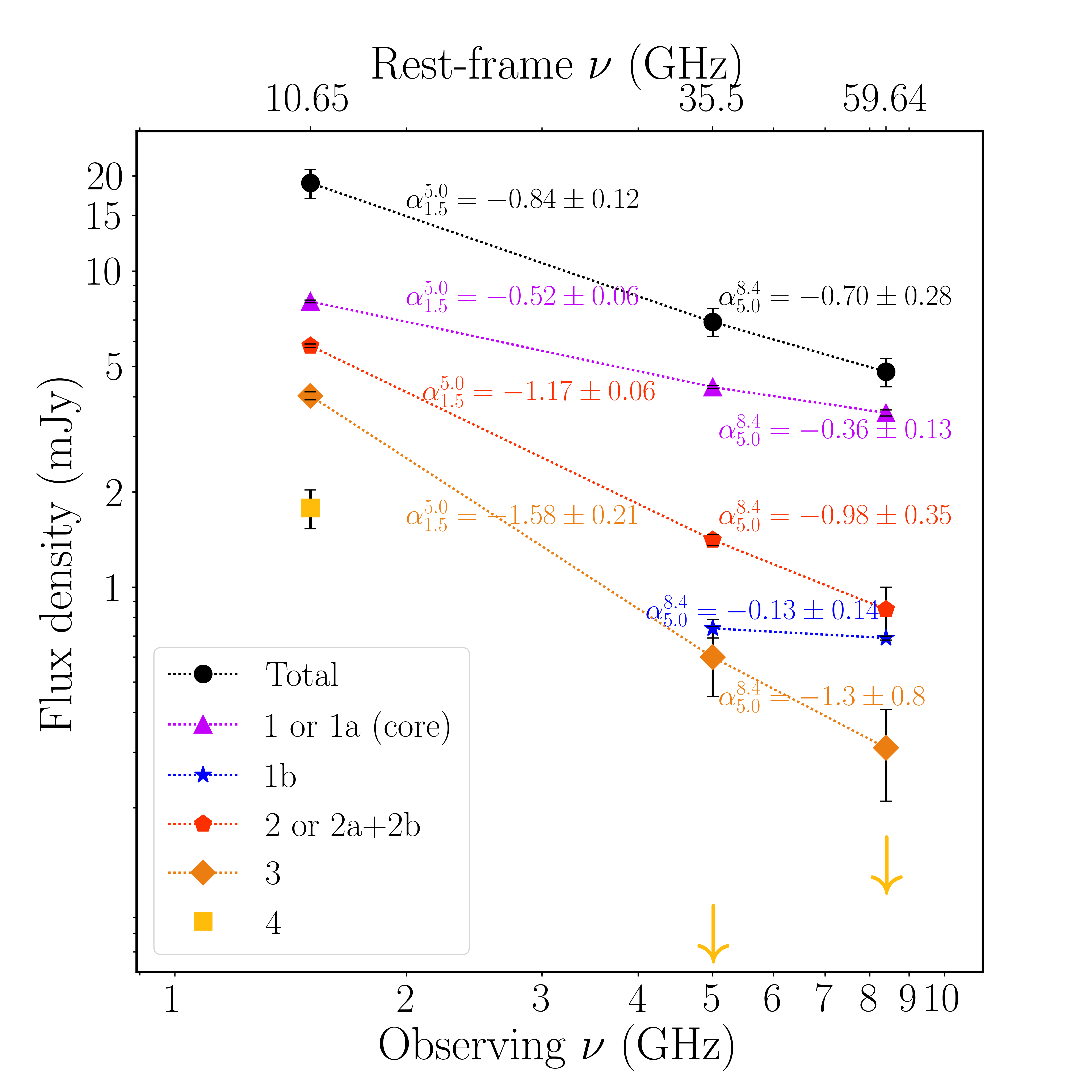}
    \caption{\textsl{Left:} Overlay of the 1.5 (black contours), 5 (white contours) and 8.4 GHz (blue contours) emission of PSO~J0309+27. The greyscale map is the 1.5 GHz self-calibrated image, and its restoring beam is shown on the bottom left corner; north is up, east is left. The phased referenced absolute position of component 1a is 03$^{\rm h}$09$^{\rm m}$47.4862$^{\rm s}$, +27$^{\circ}$17$^{\rm '}$57.3165$^{\rm ''}$. \textsl{Right:}
    Radio spectrum of all the sub-components of PSO~J0309+27 spatially resolved by the multi-frequency VLBI observations presented here. We indicate with the arrows the 3$\sigma$ detection limit for sub-component 4 at 5 and 8.4 GHz, estimated as three times the flux density within the same area of the 1.5 GHz detection of that sub-component.
    }\label{fig:VLBI_spectrum}
\end{figure*}

We fit the observed emission using 2D multi-Gaussian fits to the image-plane by using the task {\tt imfit} within the Common Astronomy Software Applications package ({\sc casa}, \citealt{McMullin2007}). The measurements of flux density, peak surface brightness and sizes obtained using this method are listed in Table~\ref{tab:vlbi_properties}, while the identified sub-components are shown in Fig.~\ref{fig:VLBI_spectrum}. In addition to the nominal uncertainties of the fit (Table~\ref{tab:vlbi_properties}), we consider the uncertainty due to the calibration process (estimated using the scatter on the amplitude gains) of the order 5 \%.

\begin{figure}
    \centering
    \includegraphics[width = 1.01\columnwidth]{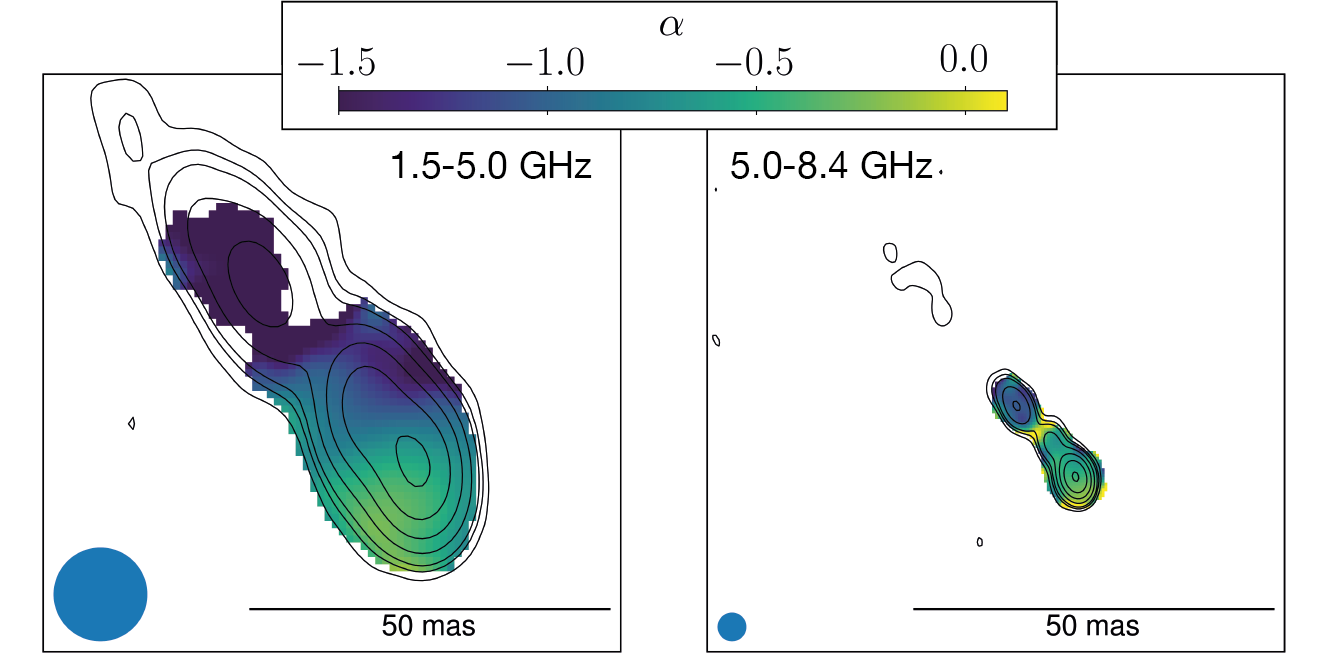}
    \caption{Spectral index maps of PSO~J0309+27 ($S_{\nu} \propto \nu^{\alpha}$) for the frequency pairs 1.5--5.0 GHz (left) and 5.0--8.4 GHz (right). The black contours are the 1.5 GHz emission on the left, and the 5 GHz emission on the right, with the same increasing scheme as Fig.~\ref{fig:VLBA_images}. The circular restoring beam is 13 mas FWHM (1.5--5.0 GHz image) and 4 mas FWHM (5.0--8.4 GHz image), and it is shown on the bottom left corner of each map.}
    \label{fig:spix_maps}
\end{figure}

We obtained synchrotron spectral index maps using images with same pixel size, convolved with the same restoring beam and matched baseline ranges. We have quantified the core-shift effect \citep{Lobanov1998} between the different frequencies by measuring the position of the core (component 1 or 1a in Table~\ref{tab:vlbi_properties}) in the CLEANed images at each band relative to the phase calibrator. This effect must be taken into account when aligning images at various frequencies for producing spectral index maps. We find that there is no positional offset between the cores at the three frequencies within the astrometric uncertainties (of the order of tens of $\mu$as). For the frequency pair 1.5--5 GHz we used the 4--35 M$\lambda$ $uv$-range and a circular restoring beam of 13 mas FWHM. For the other frequency pair 5--8.4 GHz we used visibilities in the $uv$-range of 6--100 M$\lambda$ and a circular restoring beam of 4 mas FWHM.  The resulting spectral index images are shown in Fig.~\ref{fig:spix_maps}.

\subsection{Jansky VLA}
PSO~J0309+27 was observed with the Jansky VLA (DDT observation, Legacy ID: AB1752, PI: Belladitta) in C configuration on  May 15th, 2020 from 1.4 to 40 GHz (22 to 0.9 arcsec FWHM), for a total observing time of 1.5 h (50 \% on source, 50 \% on calibration). The 1.4 and 2.3 GHz observations used the 8-bit sampler, while the other bands used the 3-bit sampler. Two antennas were excluded from the observations because of technical issues.

Calibration was done using
{\sc casa} following the standard procedure. We performed imaging (for all bands) and self-calibration (successful for L-, S-, C- and X-bands only) using the {\sc aips} package.
The 1.4 and 2.3 GHz observations were the most affected by radio frequency interferences (RFI), causing the loss of a few spectral windows (spws). We obtained separate images for each spw from 1.4 to 15 GHz using natural weights, while we averaged the 64 spws in chunks of 8 spws from 15 to 40 GHz to maximize the sensitivity. From these images we extracted the flux density of the source (unresolved at all frequencies) using a 2D Gaussian fit, and we obtained a broad-band radio spectrum from 1.4 to 40 GHz for PSO~J0309+27, which is shown in Fig.~\ref{fig:vla_spectrum}. The measurements are reported in Table~\ref{tab:appendix_vla_data}. In addition to the nominal uncertainties provided by the fit (reported in Table~\ref{tab:appendix_vla_data}), we consider the uncertainty due to the calibration process (estimated using the scatter on the amplitude gains) of the order of 3--5 \% from 1.4 to 15 GHz, while it is of 10--15 \% at 15 and 40 GHz.

\section{Results}
\label{sec:results}

\subsection{Spatially resolved multi-frequency radio emission}
\label{sec:radio_morphology}

On VLBI scales PSO~J0309+27 is a core plus one-sided jet from 1.5 to 8.4 GHz (see Fig.~\ref{fig:VLBA_images}). The jet extends at 1.5 GHz for about 80 mas at a position angle 30 degrees (east of north), which corresponds to $\sim$500~pc at the redshift of the source. At this frequency it is resolved into four steep-spectrum sub-components, which are connected by a continuous emission (Fig.~\ref{fig:VLBI_spectrum}, but see also Sec.~\ref{sec:radio_spectrum}). At 1.5 GHz we recover only $\sim$85 \% of the flux density measured by the NVSS. This can be taken as an indication for possible extended structure on scales that are resolved out by our VLBI observations. Nevertheless, this difference in flux density could be also due to intrinsic variability, as suggested by the comparison between our Jansky VLA data from May 2020 at 1.5 GHz (22 arcsec FWHM, flux density of 30 mJy) and what measured by the NVSS in 1993 (45 arcsec FWHM, flux density of about 24 mJy; see also Fig.~\ref{fig:vla_spectrum} and Sec.\ref{sec:radio_spectrum}).

At 5 and 8.4 GHz we spatially resolve the core (component 1 at 1.5 GHz) into two sub-components (1a and 1b) and the jet into three sub-components (Figs.~\ref{fig:VLBA_images} and ~\ref{fig:VLBI_spectrum}). At these frequencies we do not detect the jet sub-component 4 (clearly detected at 1.5 GHz), which may be due to its intrinsically low surface brightness and steep synchrotron spectral index. The radio core (1a) dominates the radio emission at 5 and 8.4 GHz, being about 60 \% of the total integrated flux density at both bands (see Appendix \ref{sec:appendix_tables}).

At all the three frequencies, the jet shows a shrinkage between components 2 and 3 (see Figs.~\ref{fig:VLBA_images} and \ref{fig:VLBI_spectrum}), which resembles a recollimation region right after the knot 1b \citep{Hervet2017, Bodo2018, Nishikawa2020}. Similar jet recollimations have been observed in other AGN jets in the local Universe \citep[e.g.,][]{Giroletti2012, Asada2014, Hada2018, Nakahara2019} and at high redshift \citep[e.g.,][]{Frey2015}. 

Finally, we do not find evidence for any emission that could be associated with a possible lensed image of PSO~J0309+27. Gravitationally lensing can strongly amplify the intrinsic luminosity of an AGN and, therefore, it has a crucial role in assessing the intrinsic properties of sources \citep[e.g.,][]{Spingola2020a, Spingola2020b}. We searched for possible lensed images in an area of 3 arcsec$^2$, which is the largest region not affected by bandwidth and time smearing. If gravitationally lensed, the system should more likely consist of a doubly imaged source, as there is no detection of a mirror symmetric emission that can be attributed to a merging image of a quadruply imaged lens system (Fig.~\ref{fig:VLBA_images}). By assuming a flux density ratio $\leq15:1$ and a minimum angular separation of 100 mas \citep[e.g.,][]{Spingola2019}, the second lensed image is expected to have a flux density of about 1.3 mJy at 1.5 GHz. Therefore, if present in the area, such lensed image should be detected at $30\sigma$ level.

\subsection{Radio spectrum}
\label{sec:radio_spectrum}

The spectral index is a fundamental parameter in the study of AGN jets, as it allows to clearly separate the optically thick base of the jet (the "core"), which is the closest to the SMBH, from the steep-spectrum components associated with the  more external part of the radio jets. In Fig.~\ref{fig:VLBI_spectrum} (right panel) we show the radio spectrum for all of the sub-components of PSO~J0309+27 resolved using VLBI (see left panel of Fig.~\ref{fig:VLBI_spectrum} for the labels). We can identify the brightest component (1 or 1a) as the radio core region, because it shows a flatter synchrotron spectral index between 1.5 and 5 GHz than the other sub-components ($\alpha = -0.52\pm0.06$, see Eq.~\ref{eq:sigma_spix}). Also, its spectral index becomes flatter between 5 and 8.4 GHz, as we resolve it into two sub-components (1a and 1b). Nevertheless, our 2D Gaussian fit does not associate component 1a with a point-like source, indicating that the actual source radio core is blended with the innermost optically think part of the jet. Component 1b also shows a flat spectrum between 5 and 8.4 GHz, and it is among the faintest sub-components. All of the other sub-components have steep spectral index $\alpha$, with values between $-1.58\pm0.21$ and $-0.98\pm0.35$, but the uncertainties on the integrated flux density for some of them are quite large. The most external jet component (labelled as 4) is undetected at 5 and 8.4 GHz, as described in Sec.~\ref{sec:radio_morphology}.

Spectral index maps for each frequency pair are shown in Fig.~\ref{fig:spix_maps}. There is a clear spectral index gradient between the core (flatter $\alpha$) and the jet (steeper $\alpha$). The most external jet components are clearly the steepest in PSO~J0309+27, which is consistent with what usually observed in  radio-loud AGN at all redshifts \citep{Blandford2019}. 

\begin{figure}
    \centering
    \includegraphics[width = 1.1\columnwidth]{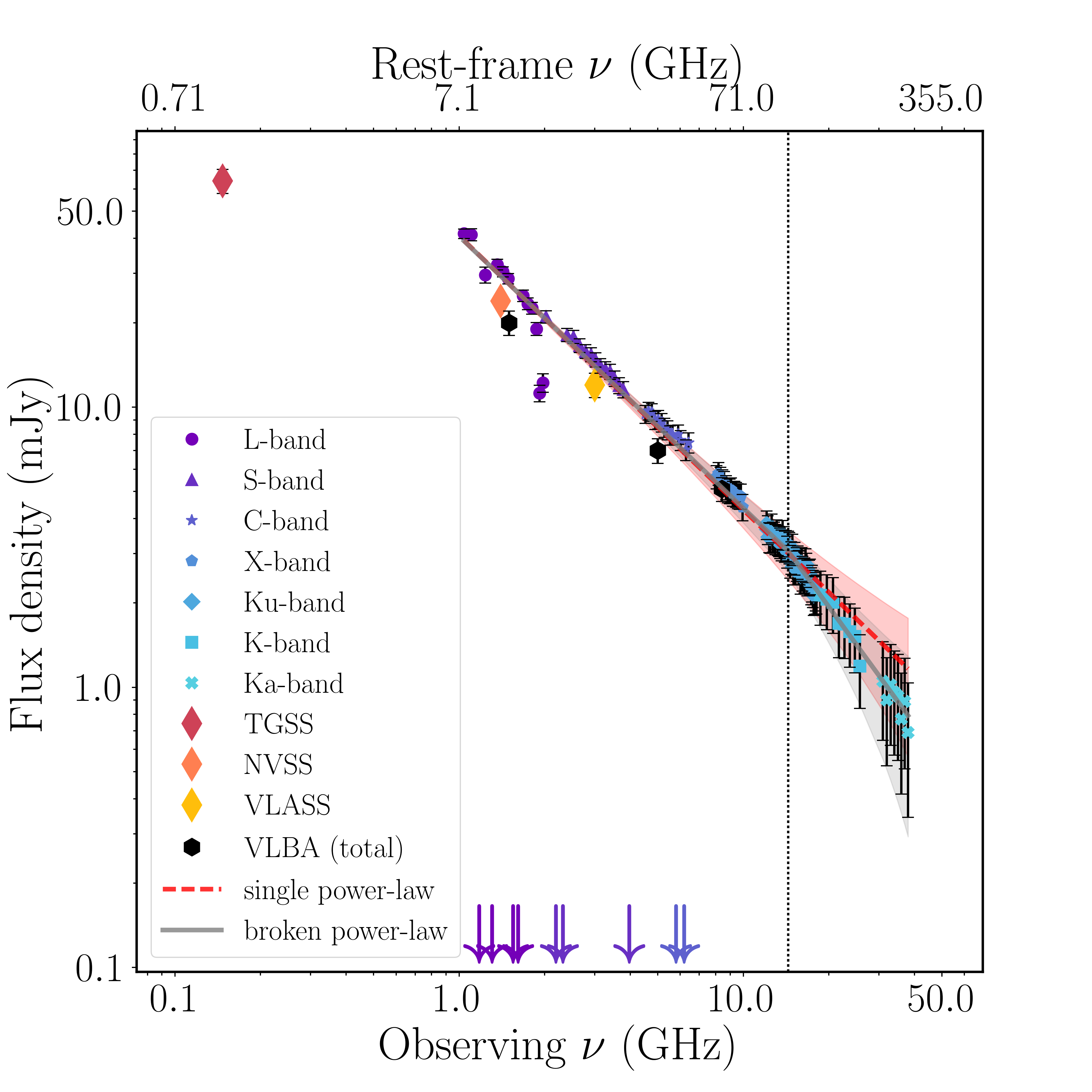}
    \caption{Radio spectrum of PSO~J0309+27 from 0.147 up to 40 GHz. The VLA observations from 1.4 to 40 GHz are represented with different symbols for each band as indicated in the legend (bottom left corner). Archival TGSS, NVSS and VLASS observations are indicated with diamonds, while the total flux density of the VLBA observations is reported using black filled circles. The arrows indicate flagged spws. The red (dashed) and grey (solid) lines stand for the single and broken power-law fits, respectively, where the shaded areas represent the uncertainties of the fits (same color coding). The vertical dotted line indicates the frequency break of the broken power-law fit (14.5 GHz).}
    \label{fig:vla_spectrum}
\end{figure}

The broad-band radio spectrum as measured by the VLA (shown in Fig.~\ref{fig:vla_spectrum}) is consistent with the steep-spectrum trend measured with the VLBA. By fitting the VLA data with a single power-law model we find a spectral index of $\alpha = -0.98 \pm 0.05$ between 1 and 40 GHz. As there is an indication for a steepening in the spectrum from the 15 GHz data (Fig.~\ref{fig:vla_spectrum}), we also fit the VLA observations using a broken power-law, which finds a break at $\nu_{\rm break} = 14.5 \pm 0.5$ GHz (rest-frame 103 GHz), a spectral index of $\alpha = -0.96\pm0.03$ before $\nu_{\rm break}$ and  $\alpha = -1.39\pm0.09$ after $\nu_{\rm break}$. 
Both the single and broken power-law models provide a good representation of the data, as they have similar reduced chi-square values (1.8 and 2.2, respectively). This is also shown in Fig.~\ref{fig:vla_spectrum}: both fits are consistent with the observations within the uncertainties.

By comparing the VLA data at 1.4 and 3 GHz with the archival observations of NVSS (at 1.4 GHz in 1993) and VLASS (at 3 GHz in 2017) we find an indication for a 20--30 \% variability in this source.  This kind of 20--30 \% variability on scales of several months to a few years\footnote{The rest-frame time difference between the NVSS and VLASS observations is $\Delta t \simeq 3.4$ years, while that between the VLASS and our VLA observations is $\Delta t \simeq 0.4$ years.} in the radio has been observed in other  radio-loud AGN, but we cannot exclude that PSO~J0309+27 had a much more pronounced variability on shorter time scales, which is typical of blazars \citep[e.g.,][]{Hovatta2008}. A future long-term monitoring is needed to assess the amplitude and the time scale of the variability in this AGN.

Moreover, we observe a flattening of the spectrum at low frequencies (between 0.147 and 1.4 GHz, Fig.~\ref{fig:vla_spectrum}). Therefore, the spectral turnover must be at low frequencies, as the flux density at 147 MHz is higher than at 1.4 GHz (Fig.~\ref{fig:vla_spectrum}).

\subsection{Physical properties }
\label{sec:physical_properties}

\begin{figure}
    \centering
    \includegraphics[width = 0.92\columnwidth]{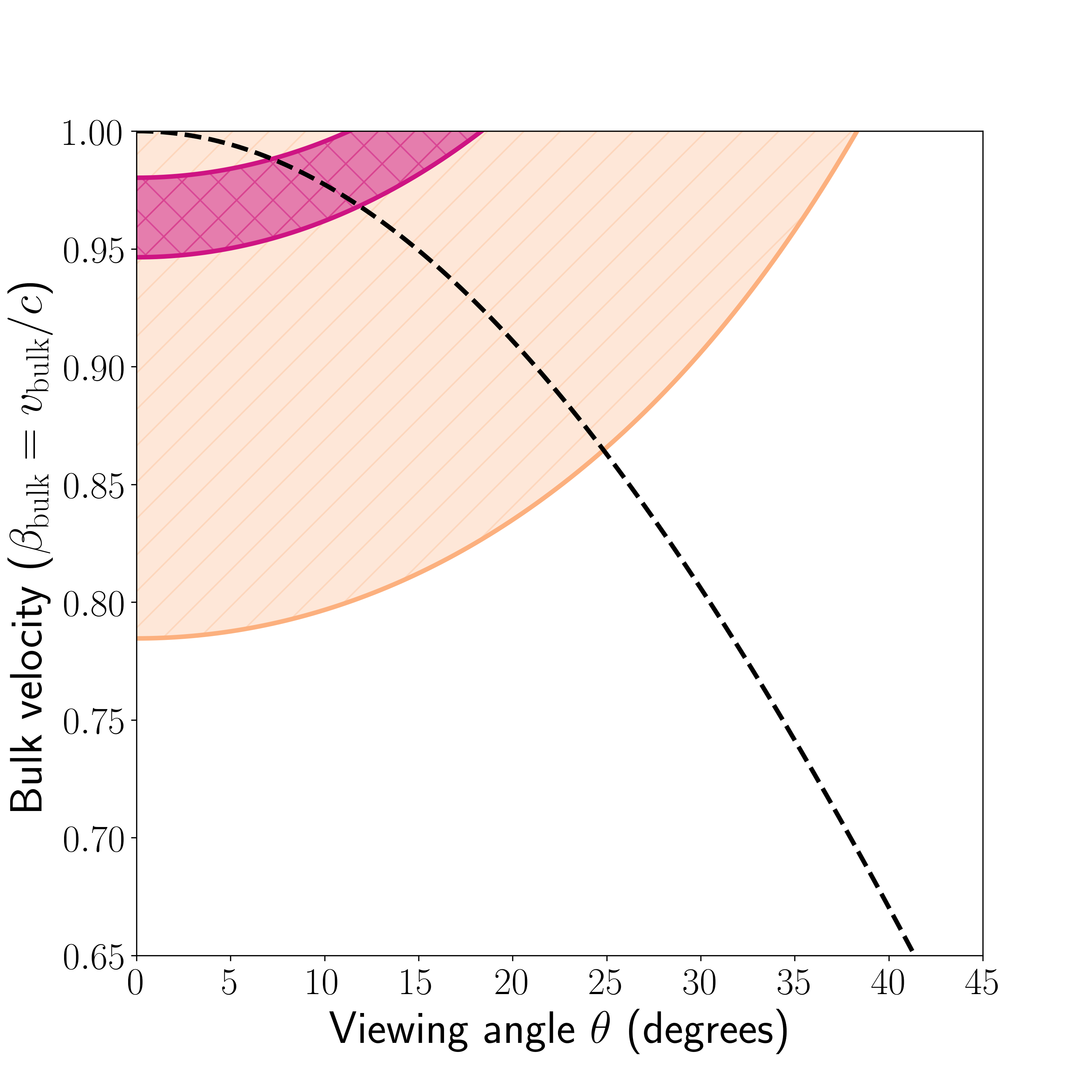}
    \caption{Allowed values on the jet viewing angle ($\theta$) and its bulk velocity ($v_{\rm bulk}$) in units of the speed of light ($\beta_{\rm bulk} = v_{\rm bulk}/c$). The orange hatched area corresponds to the allowed parameter space in the $\beta-\theta$ plane using the jet/counter-jet brightness ratio, while the magenta area traces the parameter space allowed by the correlation of \citet{Giovannini1988} as explained in Sec.~\ref{sec:physical_properties}. The black dashed line corresponds to $1/\Gamma$, which is the typical value to discern between aligned ($<1/\Gamma$) and misaligned ($>1/\Gamma$) AGN.}
    \label{fig:beta_theta}
\end{figure}

In the VLBI images we do not find any evidence for a counter-jet, which indicates that this AGN is seen under a relatively small viewing angle, as expected if PSO~J0309+27 is a blazar \citep{Belladitta2020}. To estimate the possible ranges of viewing angles $\theta$ and of the bulk velocity in terms of speed of light ($\beta_{\rm bulk}$) it is possible to use the jet/counter-jet brightness ratio $J$ (\citealt{Giovannini1994}, see Eq.~\ref{eq:jet_cj_viewing_angle}).
We use the 5 GHz observation, as it is the most sensitive among our VLBA observations, we assume the $1\sigma$ off-source noise level as an upper limit for the counter-jet surface brightness and adopt the synchrotron spectral index of component 1a between 1.5 and 5 GHz (Fig.~\ref{fig:VLBI_spectrum}).
We find a minimum bulk velocity $\beta_{\rm min}^{\rm bulk}$ of $0.78$ and a maximum viewing angle $\theta_{\rm max}$ of 38 degrees, as shown in Fig.~\ref{fig:beta_theta}. An independent way to estimate $\theta$ and the $\beta$ is given by the core dominance correlation \citep{Giovannini1988, Giovannini1994,Giroletti2004}. We used the 5 GHz flux density of the core measured from our VLBA observations (4.3 mJy) and estimated the total 408 MHz flux density by using a  spectral index of $\alpha = -0.98\pm0.05$, which is the best-fitting value for the total spectral index between 1 and 40 GHz using a single power-law model (Sec.~\ref{sec:radio_spectrum}). We account for a 30 per cent variability of the core and the scatter of the relation, finding a small range of allowed values well within the constraints from $J$ (shown in Fig.~\ref{fig:beta_theta}). As this relation has been obtained for a sample of local radio galaxies, this range should be considered only indicative and not definitive.

By assuming $\beta_{\rm min}^{\rm bulk}$ and $\theta_{\rm max}$, we find a minimum Lorentz factor $\Gamma_{\rm min} \simeq 1.59$ and a minimum Doppler factor $\delta_{\rm min} \simeq 1.62$ (see Eq.~\ref{eq:lorentz_factor} and  Eq. ~\ref{eq:doppler_factor}). These values for the lower limits on $\Gamma$ and $\delta$ are among the lowest values observed in blazars \citep[e.g.,][]{Hovatta2009, Li2020}. Nevertheless, we highlight that they are estimated by using the maximum viewing angle ($\theta_{\rm max}$) constrained by the VLBI data, which is a (shallow) upper limit on the true viewing angle of PSO~J0309+27.

We estimate the source-frame brightness temperature of each sub-component (Eq.~\ref{eq:brightness_temperature}), and the values of $T_B$ are reported in Table~\ref{tab:vlbi_properties}. The brightness temperature of the sub-components of PSO~J0309+27 is of the order of $10^8$ K at 1.5 and 5 GHz, and $10^7$ K at 8.4 GHz. We find a broad trend of decreasing $T_B$ as a function of increasing observing frequency for all of the components \citep{Lee2014}. We highlight that the rest-frame emitting frequencies are quite high, being 10.7, 35.5 and 59.6 GHz at 1.5, 5 and 8.4 GHz, respectively.

As the energy density of the CMB scales as $\propto (1+z)^4$, the inverse Compton (iC) scattering on the CMB photons (iC-CMB) is expected to play an important role at the redshift of PSO~J0309+27 \citep{Tavecchio2000,Schwartz2002,Ghisellini2014}. By using the ratio between the radio luminosity of the jet and the X-ray luminosity (Sec.~\ref{sec:appendix_formulae}, making the assumption that the X-ray emission is due to the jet only and using the radio luminosity of the jet excluding the core, Table~\ref{tab:vlbi_properties}), we find a magnetic field of $H_{\rm iC} \sim 0.09$~mG.
This assumption is supported by the high X-ray luminosity of this AGN \citep{Belladitta2020}, which highly exceeds the expectations for radio-quiet AGN \citep[e.g.,][]{Miller2011}.

This value is significantly lower that what derived by assuming equipartition conditions ($H_{\rm eq} \sim 4$~mG) and the equivalent magnetic field of the CMB at $z = 6.1$ ($H_{\rm CMB} \sim 0.17$~mG). In general, iC-CMB is more efficient than synchrotron in cooling the electrons if $H_{\rm CMB} > H_{\rm eq}$. In addition, Swift-XRT observations do not spatially resolve the X-ray emission, which may be much more extended than the VLBI jet component. Therefore, it is possible that on the X-ray scales the jet is significantly affected by iC-CMB \citep[e.g.,][]{Worrall2020}. But, as we also find that $H_{\rm eq} > H_{\rm CMB}$, we believe that  within the radio jet the energy losses are dominated by synchrotron, while the iC-CMB contribution is negligible.

\section{Discussion and conclusions}
\label{sec:discussion}

In this Letter, we report the analysis of the morphology and physical properties of PSO~J0309+27 spatially resolved on parsec scales for the first time by our DDT follow-up radio observations. With the sensitive VLBA observations we unveil a bright jet extended on $\sim500$ parsecs in projection, which is resolved into multiple sub-components  (Fig.~\ref{fig:VLBA_images}).

We find brightness temperature values around $\sim10^8$ K. Similar values for $T_B$ associated with bright extended steep-spectrum jets (from a few to hundreds of parsecs in projection) has been measured also in other radio-loud AGN at redshifts $z>4$ (\citealt{Frey2011,Gabanyi2015, Cao2017,Banados2018b, Zhu2018,Momjian2018,An2020}, but see \citealt{Frey2015} for higher $T_B$ values at $z\sim5$). These $T_B$ estimates are 2-3 order of magnitudes lower than what measured for the blazar sample of MOJAVE observed at 15 GHz 
(\citealt{Kovalev2005, Homan2006}; at redshifts up to $z \simeq3.5$, \citealt{Cara2008}), and the VLBA-BU-BLAZAR sample observed at 43 GHz (at redshifts up to $z \simeq2.5$, \citealt{Jorstad2016,Jorstad2017}).
Also, the frequency-dependency of $T_B$ \citep{Lee2014}, the turnover of the synchrotron emission being in the MHz regime (rest-frame $1-7$ GHz, Fig.~\ref{fig:vla_spectrum}) and the very high rest-frame frequencies sampled by the observations ($\sim 7-300$ GHz,  Fig.~\ref{fig:vla_spectrum}) presented here naturally lead to lower values of $T_B$ in PSO~J0309+27.

Another issue in the estimate of the Doppler factor and $T_B$ is related to the single epoch measurement. Such estimates from a single epoch observation are usually much lower than what derived from variability (e.g., \citealt{Orienti2015}). This is because variability may overcome the intrinsic limitations due to the angular resolution of the instrument, and can probe much smaller emitting regions simply because of causality arguments.

Our derived values for the magnetic field in PSO~J0309+27 indicate that iC-CMB process is negligible within the radio jet detected on VLBI-scales, while it could be important for the X--ray emission.

Overall, the physical properties of this AGN at redshift $z=6.1$ point to a moderately beamed jet (Sec.~\ref{sec:physical_properties} and Table~\ref{tab:vlbi_properties}). However, our observational constraints deriving from a single epoch observation are too loose to obtain stringent values for parameters like the bulk velocity and the viewing angle of the jet, so that we could mainly put upper (or lower) limits. Nevertheless, if PSO~J0309+27 is a blazar, as suggested by its X-ray properties \citep{Belladitta2020}, then in Fig.~\ref{fig:beta_theta}  it should lie below the dashed line, which is the typical value to discern between aligned ($<1/\Gamma$) and misaligned ($>1/\Gamma$) AGN, and below the maximum value allowed by the core-dominance relation (that is $\beta_{\rm bulk} = 0.98$). This implies a Lorentz factor $\Gamma$ of $\simeq 5$ or less (using Eq.~\ref{eq:lorentz_factor}). A value of $\Gamma$ of $\simeq 5$ would be in very good agreement with what proposed by \citet{Volonteri2011} to reconcile the significant deficit of observed high-z blazars with respect to the expected number of blazars at $z\gtrapprox 3$. 

Finally, we highlight that our findings cannot exclude that PSO~J0309+27 is seen under a larger viewing angle. In such case, the values for $\Gamma$ could be higher, but the bright X-ray emission would need a boost, which could be due to iC-CMB.
More stringent constraints on the bulk Lorentz factor in PSO~J0309+27 and the other high-z blazars are necessary to test whether $\Gamma$ is intrinsically low in distant blazars and actually changes as a function of redshift. 

\begin{acknowledgements}
 We thank the anonymous referee for their constructive comments to the manuscript.
CS thanks the NRAO analysts for their prompt help with the preparation of the schedule and delivery of the data during the critical period of April--May 2020 due to the COVID-19 pandemic. The authors are grateful to Leonid Gurvits and Alexey Melnikov for providing information on the Russian VLBI-network “Quasar” observations of this target and the useful discussions on this work.

CS is grateful for support from the National Research Council of Science and Technology, Korea (EU-16-001).

The National Radio Astronomy Observatory is a facility of the National Science Foundation operated under cooperative agreement by Associated Universities, Inc.
  
This research made use of APLpy, an open-source plotting package for Python \citep{Robitaille2012}.
\end{acknowledgements}

\bibliographystyle{aa} 
\bibpunct{(}{)}{;}{a}{}{,} 
\bibliography{references} 

\begin{thebibliography}{71}
\expandafter\ifx\csname natexlab\endcsname\relax\def\natexlab#1{#1}\fi

\bibitem[{{Ajello} {et~al.}(2009){Ajello}, {Costamante}, {Sambruna}, {Gehrels},
  {Chiang}, {Rau}, {Escala}, {Greiner}, {Tueller}, {Wall}, \&
  {Mushotzky}}]{Ajello2009}
{Ajello}, M., {Costamante}, L., {Sambruna}, R.~M., {et~al.} 2009, \apj, 699,
  603

\bibitem[{{An} {et~al.}(2020){An}, {Mohan}, {Zhang}, {Frey}, {Yang},
  {Gab{\'a}nyi}, {Gurvits}, {Paragi}, {Perger}, \& {Zheng}}]{An2020}
{An}, T., {Mohan}, P., {Zhang}, Y., {et~al.} 2020, Nature Communications, 11,
  143

\bibitem[{{Asada} {et~al.}(2014){Asada}, {Nakamura}, {Doi}, {Nagai}, \&
  {Inoue}}]{Asada2014}
{Asada}, K., {Nakamura}, M., {Doi}, A., {Nagai}, H., \& {Inoue}, M. 2014,
  \apjl, 781, L2

\bibitem[{{Ba{\~n}ados} {et~al.}(2018){Ba{\~n}ados}, {Carilli}, {Walter},
  {Momjian}, {Decarli}, {Farina}, {Mazzucchelli}, \& {Venemans}}]{Banados2018b}
{Ba{\~n}ados}, E., {Carilli}, C., {Walter}, F., {et~al.} 2018, \apjl, 861, L14

\bibitem[{{Ba{\~n}ados} {et~al.}(2015){Ba{\~n}ados}, {Venemans}, {Morganson},
  {Hodge}, {Decarli}, {Walter}, {Stern}, {Schlafly}, {Farina}, {Greiner},
  {Chambers}, {Fan}, {Rix}, {Burgett}, {Draper}, {Flewelling}, {Kaiser},
  {Metcalfe}, {Morgan}, {Tonry}, \& {Wainscoat}}]{Banados2015}
{Ba{\~n}ados}, E., {Venemans}, B.~P., {Morganson}, E., {et~al.} 2015, \apj,
  804, 118

\bibitem[{{Belladitta} {et~al.}(2020){Belladitta}, {Moretti}, {Caccianiga},
  {Spingola}, {Severgnini}, {Della Ceca}, {Ghisellini}, {Dallacasa},
  {Sbarrato}, {Cicone}, {Cassar{\`a}}, \& {Pedani}}]{Belladitta2020}
{Belladitta}, S., {Moretti}, A., {Caccianiga}, A., {et~al.} 2020, \aap, 635, L7

\bibitem[{{Blandford} {et~al.}(2019){Blandford}, {Meier}, \&
  {Readhead}}]{Blandford2019}
{Blandford}, R., {Meier}, D., \& {Readhead}, A. 2019, \araa, 57, 467

\bibitem[{{Bodo} \& {Tavecchio}(2018)}]{Bodo2018}
{Bodo}, G. \& {Tavecchio}, F. 2018, \aap, 609, A122

\bibitem[{{Caccianiga} {et~al.}(2019){Caccianiga}, {Moretti}, {Belladitta},
  {Della Ceca}, {Ant{\'o}n}, {Ballo}, {Cicone}, {Dallacasa}, {Gargiulo},
  {Ighina}, {March{\~a}}, \& {Severgnini}}]{Caccianiga2019}
{Caccianiga}, A., {Moretti}, A., {Belladitta}, S., {et~al.} 2019, \mnras, 484,
  204

\bibitem[{{Cao} {et~al.}(2017){Cao}, {Frey}, {Gab{\'a}nyi}, {Paragi}, {Yang},
  {Cseh}, {Hong}, \& {An}}]{Cao2017}
{Cao}, H.~M., {Frey}, S., {Gab{\'a}nyi}, K.~{\'E}., {et~al.} 2017, \mnras, 467,
  950

\bibitem[{{Cara} \& {Lister}(2008)}]{Cara2008}
{Cara}, M. \& {Lister}, M.~L. 2008, \apj, 674, 111

\bibitem[{{Chambers} {et~al.}(2016){Chambers}, {Magnier}, {Metcalfe},
  {Flewelling}, {Huber}, {Waters}, {Denneau}, {Draper}, {Farrow}, {Finkbeiner},
  {Holmberg}, {Koppenhoefer}, {Price}, {Rest}, {Saglia}, {Schlafly}, {Smartt},
  {Sweeney}, {Wainscoat}, {Burgett}, {Chastel}, {Grav}, {Heasley}, {Hodapp},
  {Jedicke}, {Kaiser}, {Kudritzki}, {Luppino}, {Lupton}, {Monet}, {Morgan},
  {Onaka}, {Shiao}, {Stubbs}, {Tonry}, {White}, {Ba{\~n}ados}, {Bell},
  {Bender}, {Bernard}, {Boegner}, {Boffi}, {Botticella}, {Calamida},
  {Casertano}, {Chen}, {Chen}, {Cole}, {Deacon}, {Frenk}, {Fitzsimmons},
  {Gezari}, {Gibbs}, {Goessl}, {Goggia}, {Gourgue}, {Goldman}, {Grant},
  {Grebel}, {Hambly}, {Hasinger}, {Heavens}, {Heckman}, {Henderson}, {Henning},
  {Holman}, {Hopp}, {Ip}, {Isani}, {Jackson}, {Keyes}, {Koekemoer}, {Kotak},
  {Le}, {Liska}, {Long}, {Lucey}, {Liu}, {Martin}, {Masci}, {McLean}, {Mindel},
  {Misra}, {Morganson}, {Murphy}, {Obaika}, {Narayan}, {Nieto-Santisteban},
  {Norberg}, {Peacock}, {Pier}, {Postman}, {Primak}, {Rae}, {Rai}, {Riess},
  {Riffeser}, {Rix}, {R{\"o}ser}, {Russel}, {Rutz}, {Schilbach}, {Schultz},
  {Scolnic}, {Strolger}, {Szalay}, {Seitz}, {Small}, {Smith}, {Soderblom},
  {Taylor}, {Thomson}, {Taylor}, {Thakar}, {Thiel}, {Thilker}, {Unger},
  {Urata}, {Valenti}, {Wagner}, {Walder}, {Walter}, {Watters}, {Werner},
  {Wood-Vasey}, \& {Wyse}}]{Chambers2016}
{Chambers}, K.~C., {Magnier}, E.~A., {Metcalfe}, N., {et~al.} 2016, arXiv
  e-prints, arXiv:1612.05560

\bibitem[{{Condon} {et~al.}(1998){Condon}, {Cotton}, {Greisen}, {Yin},
  {Perley}, {Taylor}, \& {Broderick}}]{Condon1998}
{Condon}, J.~J., {Cotton}, W.~D., {Greisen}, E.~W., {et~al.} 1998, \aj, 115,
  1693

\bibitem[{{Deller} {et~al.}(2011){Deller}, {Brisken}, {Phillips}, {Morgan},
  {Alef}, {Cappallo}, {Middelberg}, {Romney}, {Rottmann}, {Tingay}, \&
  {Wayth}}]{Deller2011}
{Deller}, A.~T., {Brisken}, W.~F., {Phillips}, C.~J., {et~al.} 2011, \pasp,
  123, 275

\bibitem[{{Fabian}(2012)}]{Fabian2012}
{Fabian}, A.~C. 2012, \araa, 50, 455

\bibitem[{{Fixsen}(2009)}]{Fixsen2009}
{Fixsen}, D.~J. 2009, \apj, 707, 916

\bibitem[{{Frey} {et~al.}(2015){Frey}, {Paragi}, {Fogasy}, \&
  {Gurvits}}]{Frey2015}
{Frey}, S., {Paragi}, Z., {Fogasy}, J.~O., \& {Gurvits}, L.~I. 2015, \mnras,
  446, 2921

\bibitem[{{Frey} {et~al.}(2011){Frey}, {Paragi}, {Gurvits}, {Gab{\'a}nyi}, \&
  {Cseh}}]{Frey2011}
{Frey}, S., {Paragi}, Z., {Gurvits}, L.~I., {Gab{\'a}nyi}, K.~{\'E}., \&
  {Cseh}, D. 2011, \aap, 531, L5

\bibitem[{{Gabanyi} {et~al.}(2015){Gabanyi}, {Cseh}, {Frey}, {Paragi},
  {Gurvits}, {An}, \& {Zhang}}]{Gabanyi2015}
{Gabanyi}, K.~E., {Cseh}, D., {Frey}, S., {et~al.} 2015, \mnras, 450, L57

\bibitem[{{Ghisellini} {et~al.}(2014){Ghisellini}, {Celotti}, {Tavecchio},
  {Haardt}, \& {Sbarrato}}]{Ghisellini2014}
{Ghisellini}, G., {Celotti}, A., {Tavecchio}, F., {Haardt}, F., \& {Sbarrato},
  T. 2014, \mnras, 438, 2694

\bibitem[{{Giovannini} {et~al.}(1988){Giovannini}, {Feretti}, {Gregorini}, \&
  {Parma}}]{Giovannini1988}
{Giovannini}, G., {Feretti}, L., {Gregorini}, L., \& {Parma}, P. 1988, \aap,
  199, 73

\bibitem[{{Giovannini} {et~al.}(1994){Giovannini}, {Feretti}, {Venturi},
  {Lara}, {Marcaide}, {Rioja}, {Spangler}, \& {Wehrle}}]{Giovannini1994}
{Giovannini}, G., {Feretti}, L., {Venturi}, T., {et~al.} 1994, \apj, 435, 116

\bibitem[{{Giroletti} {et~al.}(2004){Giroletti}, {Giovannini}, {Feretti},
  {Cotton}, {Edwards}, {Lara}, {Marscher}, {Mattox}, {Piner}, \&
  {Venturi}}]{Giroletti2004}
{Giroletti}, M., {Giovannini}, G., {Feretti}, L., {et~al.} 2004, \apj, 600, 127

\bibitem[{{Giroletti} {et~al.}(2012){Giroletti}, {Hada}, {Giovannini},
  {Casadio}, {Beilicke}, {Cesarini}, {Cheung}, {Doi}, {Krawczynski}, {Kino},
  {Lee}, \& {Nagai}}]{Giroletti2012}
{Giroletti}, M., {Hada}, K., {Giovannini}, G., {et~al.} 2012, \aap, 538, L10

\bibitem[{{Greisen}(2003)}]{Greisen2003}
{Greisen}, E.~W. 2003, Astrophysics and Space Science Library, Vol. 285, {AIPS,
  the VLA, and the VLBA}, ed. A.~{Heck}, 109

\bibitem[{{Gurvits} {et~al.}(1999){Gurvits}, {Kellermann}, \&
  {Frey}}]{Gurvits1999}
{Gurvits}, L.~I., {Kellermann}, K.~I., \& {Frey}, S. 1999, \aap, 342, 378

\bibitem[{{Hada} {et~al.}(2018){Hada}, {Doi}, {Wajima}, {D'Ammand o},
  {Orienti}, {Giroletti}, {Giovannini}, {Nakamura}, \& {Asada}}]{Hada2018}
{Hada}, K., {Doi}, A., {Wajima}, K., {et~al.} 2018, \apj, 860, 141

\bibitem[{{Haiman} {et~al.}(2004){Haiman}, {Quataert}, \& {Bower}}]{Haiman2004}
{Haiman}, Z., {Quataert}, E., \& {Bower}, G.~C. 2004, \apj, 612, 698

\bibitem[{{Hervet} {et~al.}(2017){Hervet}, {Meliani}, {Zech}, {Boisson},
  {Cayatte}, {Sauty}, \& {Sol}}]{Hervet2017}
{Hervet}, O., {Meliani}, Z., {Zech}, A., {et~al.} 2017, \aap, 606, A103

\bibitem[{{Homan} {et~al.}(2006){Homan}, {Kovalev}, {Lister}, {Ros},
  {Kellermann}, {Cohen}, {Vermeulen}, {Zensus}, \& {Kadler}}]{Homan2006}
{Homan}, D.~C., {Kovalev}, Y.~Y., {Lister}, M.~L., {et~al.} 2006, \apjl, 642,
  L115

\bibitem[{{Hovatta} {et~al.}(2008){Hovatta}, {Nieppola}, {Tornikoski},
  {Valtaoja}, {Aller}, \& {Aller}}]{Hovatta2008}
{Hovatta}, T., {Nieppola}, E., {Tornikoski}, M., {et~al.} 2008, \aap, 485, 51

\bibitem[{{Hovatta} {et~al.}(2009){Hovatta}, {Valtaoja}, {Tornikoski}, \&
  {L{\"a}hteenm{\"a}ki}}]{Hovatta2009}
{Hovatta}, T., {Valtaoja}, E., {Tornikoski}, M., \& {L{\"a}hteenm{\"a}ki}, A.
  2009, \aap, 494, 527

\bibitem[{{Intema} {et~al.}(2017){Intema}, {Jagannathan}, {Mooley}, \&
  {Frail}}]{Intema2017}
{Intema}, H.~T., {Jagannathan}, P., {Mooley}, K.~P., \& {Frail}, D.~A. 2017,
  \aap, 598, A78

\bibitem[{{Jorstad} \& {Marscher}(2016)}]{Jorstad2016}
{Jorstad}, S. \& {Marscher}, A. 2016, Galaxies, 4, 47

\bibitem[{{Jorstad} {et~al.}(2017){Jorstad}, {Marscher}, {Morozova},
  {Troitsky}, {Agudo}, {Casadio}, {Foord}, {G{\'o}mez}, {MacDonald}, {Molina},
  {L{\"a}hteenm{\"a}ki}, {Tammi}, \& {Tornikoski}}]{Jorstad2017}
{Jorstad}, S.~G., {Marscher}, A.~P., {Morozova}, D.~A., {et~al.} 2017, \apj,
  846, 98

\bibitem[{{Kovalev} {et~al.}(2005){Kovalev}, {Kellermann}, {Lister}, {Homan},
  {Vermeulen}, {Cohen}, {Ros}, {Kadler}, {Lobanov}, {Zensus}, {Kardashev},
  {Gurvits}, {Aller}, \& {Aller}}]{Kovalev2005}
{Kovalev}, Y.~Y., {Kellermann}, K.~I., {Lister}, M.~L., {et~al.} 2005, \aj,
  130, 2473

\bibitem[{{Lacy} {et~al.}(2019){Lacy}, {Chandler}, {Kimball}, {Myers},
  {Nyland}, \& {Witz}}]{Lacy2019}
{Lacy}, M., {Chandler}, C., {Kimball}, A., {et~al.} 2019, in Astronomical
  Society of the Pacific Conference Series, Vol. 523, Astronomical Data
  Analysis Software and Systems XXVII, ed. P.~J. {Teuben}, M.~W. {Pound}, B.~A.
  {Thomas}, \& E.~M. {Warner}, 217

\bibitem[{{Lee}(2014)}]{Lee2014}
{Lee}, S.-S. 2014, Journal of Korean Astronomical Society, 47, 303

\bibitem[{{Li} {et~al.}(2020){Li}, {An}, {Mohan}, \& {Giroletti}}]{Li2020}
{Li}, X., {An}, T., {Mohan}, P., \& {Giroletti}, M. 2020, arXiv e-prints,
  arXiv:2005.00300

\bibitem[{{Lister} {et~al.}(2013){Lister}, {Aller}, {Aller}, {Homan},
  {Kellermann}, {Kovalev}, {Pushkarev}, {Richards}, {Ros}, \&
  {Savolainen}}]{Lister2013}
{Lister}, M.~L., {Aller}, M.~F., {Aller}, H.~D., {et~al.} 2013, \aj, 146, 120

\bibitem[{{Lobanov}(2010)}]{Lobanov2010}
{Lobanov}, A. 2010, arXiv e-prints, arXiv:1010.2856

\bibitem[{{Lobanov}(1998)}]{Lobanov1998}
{Lobanov}, A.~P. 1998, \aaps, 132, 261

\bibitem[{{Mainzer} {et~al.}(2011){Mainzer}, {Grav}, {Masiero}, {Bauer},
  {Cutri}, {McMillan}, {Wright}, \& {Spahr}}]{Mainzer2011}
{Mainzer}, A., {Grav}, T., {Masiero}, J., {et~al.} 2011, in EPSC-DPS Joint
  Meeting 2011, Vol. 2011, 1530

\bibitem[{{McMullin} {et~al.}(2007){McMullin}, {Waters}, {Schiebel}, {Young},
  \& {Golap}}]{McMullin2007}
{McMullin}, J.~P., {Waters}, B., {Schiebel}, D., {Young}, W., \& {Golap}, K.
  2007, in Astronomical Society of the Pacific Conference Series, Vol. 376,
  Astronomical Data Analysis Software and Systems XVI, ed. R.~A. {Shaw},
  F.~{Hill}, \& D.~J. {Bell}, 127

\bibitem[{{Medvedev} {et~al.}(2020){Medvedev}, {Sazonov}, {Gilfanov},
  {Burenin}, {Khorunzhev}, {Meshcheryakov}, {Sunyaev}, {Bikmaev}, \&
  {Irtuganov}}]{Medvedev2020}
{Medvedev}, P., {Sazonov}, S., {Gilfanov}, M., {et~al.} 2020, arXiv e-prints,
  arXiv:2007.04735

\bibitem[{{Miller} {et~al.}(2011){Miller}, {Brandt}, {Schneider}, {Gibson},
  {Steffen}, \& {Wu}}]{Miller2011}
{Miller}, B.~P., {Brandt}, W.~N., {Schneider}, D.~P., {et~al.} 2011, \apj, 726,
  20

\bibitem[{{Momjian} {et~al.}(2018){Momjian}, {Carilli}, {Ba{\~n}ados},
  {Walter}, \& {Venemans}}]{Momjian2018}
{Momjian}, E., {Carilli}, C.~L., {Ba{\~n}ados}, E., {Walter}, F., \&
  {Venemans}, B.~P. 2018, \apj, 861, 86

\bibitem[{{Nakahara} {et~al.}(2019){Nakahara}, {Doi}, {Murata}, {Nakamura},
  {Hada}, \& {Asada}}]{Nakahara2019}
{Nakahara}, S., {Doi}, A., {Murata}, Y., {et~al.} 2019, \apj, 878, 61

\bibitem[{{Nishikawa} {et~al.}(2020){Nishikawa}, {Mizuno}, {G{\'o}mez},
  {Du{\c{t}}an}, {Niemiec}, {Kobzar}, {MacDonald}, {Meli}, {Pohl}, \&
  {Hirotani}}]{Nishikawa2020}
{Nishikawa}, K., {Mizuno}, Y., {G{\'o}mez}, J.~L., {et~al.} 2020, \mnras, 493,
  2652

\bibitem[{{Orienti} {et~al.}(2015){Orienti}, {D'Ammando}, {Larsson}, {Finke},
  {Giroletti}, {Dallacasa}, {Isacsson}, \& {Stoby Hoglund}}]{Orienti2015}
{Orienti}, M., {D'Ammando}, F., {Larsson}, J., {et~al.} 2015, \mnras, 453, 4037

\bibitem[{{Pacholczyk}(1970)}]{Pacholczyk1970}
{Pacholczyk}, A.~G. 1970, {Radio astrophysics. Nonthermal processes in galactic
  and extragalactic sources}

\bibitem[{{Padovani} {et~al.}(2017){Padovani}, {Alexander}, {Assef}, {De
  Marco}, {Giommi}, {Hickox}, {Richards}, {Smol{\v{c}}i{\'c}},
  {Hatziminaoglou}, {Mainieri}, \& {Salvato}}]{Padovani2017}
{Padovani}, P., {Alexander}, D.~M., {Assef}, R.~J., {et~al.} 2017, \aapr, 25, 2

\bibitem[{{Padovani} {et~al.}(2015){Padovani}, {Bonzini}, {Kellermann},
  {Miller}, {Mainieri}, \& {Tozzi}}]{Padovani2015}
{Padovani}, P., {Bonzini}, M., {Kellermann}, K.~I., {et~al.} 2015, \mnras, 452,
  1263

\bibitem[{{Planck Collaboration} {et~al.}(2016){Planck Collaboration}, {Ade},
  {Aghanim}, {Arnaud}, {Ashdown}, {Aumont}, {Baccigalupi}, {Banday},
  {Barreiro}, {Bartlett}, {Bartolo}, {Battaner}, {Battye}, {Benabed},
  {Beno{\^\i}t}, {Benoit-L{\'e}vy}, {Bernard}, {Bersanelli}, {Bielewicz},
  {Bock}, {Bonaldi}, {Bonavera}, {Bond}, {Borrill}, {Bouchet}, {Boulanger},
  {Bucher}, {Burigana}, {Butler}, {Calabrese}, {Cardoso}, {Catalano},
  {Challinor}, {Chamballu}, {Chary}, {Chiang}, {Chluba}, {Christensen},
  {Church}, {Clements}, {Colombi}, {Colombo}, {Combet}, {Coulais}, {Crill},
  {Curto}, {Cuttaia}, {Danese}, {Davies}, {Davis}, {de Bernardis}, {de Rosa},
  {de Zotti}, {Delabrouille}, {D{\'e}sert}, {Di Valentino}, {Dickinson},
  {Diego}, {Dolag}, {Dole}, {Donzelli}, {Dor{\'e}}, {Douspis}, {Ducout},
  {Dunkley}, {Dupac}, {Efstathiou}, {Elsner}, {En{\ss}lin}, {Eriksen},
  {Farhang}, {Fergusson}, {Finelli}, {Forni}, {Frailis}, {Fraisse},
  {Franceschi}, {Frejsel}, {Galeotta}, {Galli}, {Ganga}, {Gauthier}, {Gerbino},
  {Ghosh}, {Giard}, {Giraud-H{\'e}raud}, {Giusarma}, {Gjerl{\o}w},
  {Gonz{\'a}lez-Nuevo}, {G{\'o}rski}, {Gratton}, {Gregorio}, {Gruppuso},
  {Gudmundsson}, {Hamann}, {Hansen}, {Hanson}, {Harrison}, {Helou},
  {Henrot-Versill{\'e}}, {Hern{\'a}ndez-Monteagudo}, {Herranz}, {Hildebrand t},
  {Hivon}, {Hobson}, {Holmes}, {Hornstrup}, {Hovest}, {Huang}, {Huffenberger},
  {Hurier}, {Jaffe}, {Jaffe}, {Jones}, {Juvela}, {Keih{\"a}nen}, {Keskitalo},
  {Kisner}, {Kneissl}, {Knoche}, {Knox}, {Kunz}, {Kurki-Suonio}, {Lagache},
  {L{\"a}hteenm{\"a}ki}, {Lamarre}, {Lasenby}, {Lattanzi}, {Lawrence}, {Leahy},
  {Leonardi}, {Lesgourgues}, {Levrier}, {Lewis}, {Liguori}, {Lilje},
  {Linden-V{\o}rnle}, {L{\'o}pez-Caniego}, {Lubin}, {Mac{\'\i}as-P{\'e}rez},
  {Maggio}, {Maino}, {Mandolesi}, {Mangilli}, {Marchini}, {Maris}, {Martin},
  {Martinelli}, {Mart{\'\i}nez-Gonz{\'a}lez}, {Masi}, {Matarrese}, {McGehee},
  {Meinhold}, {Melchiorri}, {Melin}, {Mendes}, {Mennella}, {Migliaccio},
  {Millea}, {Mitra}, {Miville-Desch{\^e}nes}, {Moneti}, {Montier}, {Morgante},
  {Mortlock}, {Moss}, {Munshi}, {Murphy}, {Naselsky}, {Nati}, {Natoli},
  {Netterfield}, {N{\o}rgaard-Nielsen}, {Noviello}, {Novikov}, {Novikov},
  {Oxborrow}, {Paci}, {Pagano}, {Pajot}, {Paladini}, {Paoletti}, {Partridge},
  {Pasian}, {Patanchon}, {Pearson}, {Perdereau}, {Perotto}, {Perrotta},
  {Pettorino}, {Piacentini}, {Piat}, {Pierpaoli}, {Pietrobon}, {Plaszczynski},
  {Pointecouteau}, {Polenta}, {Popa}, {Pratt}, {Pr{\'e}zeau}, {Prunet},
  {Puget}, {Rachen}, {Reach}, {Rebolo}, {Reinecke}, {Remazeilles}, {Renault},
  {Renzi}, {Ristorcelli}, {Rocha}, {Rosset}, {Rossetti}, {Roudier},
  {Rouill{\'e} d'Orfeuil}, {Rowan-Robinson}, {Rubi{\~n}o-Mart{\'\i}n},
  {Rusholme}, {Said}, {Salvatelli}, {Salvati}, {Sandri}, {Santos},
  {Savelainen}, {Savini}, {Scott}, {Seiffert}, {Serra}, {Shellard}, {Spencer},
  {Spinelli}, {Stolyarov}, {Stompor}, {Sudiwala}, {Sunyaev}, {Sutton},
  {Suur-Uski}, {Sygnet}, {Tauber}, {Terenzi}, {Toffolatti}, {Tomasi},
  {Tristram}, {Trombetti}, {Tucci}, {Tuovinen}, {T{\"u}rler}, {Umana},
  {Valenziano}, {Valiviita}, {Van Tent}, {Vielva}, {Villa}, {Wade}, {Wandelt},
  {Wehus}, {White}, {White}, {Wilkinson}, {Yvon}, {Zacchei}, \&
  {Zonca}}]{Planck2016}
{Planck Collaboration}, {Ade}, P.~A.~R., {Aghanim}, N., {et~al.} 2016, \aap,
  594, A13

\bibitem[{{Robitaille} \& {Bressert}(2012)}]{Robitaille2012}
{Robitaille}, T. \& {Bressert}, E. 2012, {APLpy: Astronomical Plotting Library
  in Python}

\bibitem[{{Romani} {et~al.}(2004){Romani}, {Sowards-Emmerd}, {Greenhill}, \&
  {Michelson}}]{Romani2004}
{Romani}, R.~W., {Sowards-Emmerd}, D., {Greenhill}, L., \& {Michelson}, P.
  2004, \apjl, 610, L9

\bibitem[{{Schwartz}(2002)}]{Schwartz2002}
{Schwartz}, D.~A. 2002, \apjl, 571, L71

\bibitem[{{Spingola} \& {Barnacka}(2020)}]{Spingola2020a}
{Spingola}, C. \& {Barnacka}, A. 2020, \mnras, 494, 2312

\bibitem[{{Spingola} {et~al.}(2019){Spingola}, {McKean}, {Lee}, {Deller}, \&
  {Moldon}}]{Spingola2019}
{Spingola}, C., {McKean}, J.~P., {Lee}, M., {Deller}, A., \& {Moldon}, J. 2019,
  \mnras, 483, 2125

\bibitem[{{Spingola} {et~al.}(2020){Spingola}, {McKean}, {Vegetti}, {Powell},
  {Auger}, {Koopmans}, {Fassnacht}, {Lagattuta}, {Rizzo}, {Stacey}, \&
  {Sweijen}}]{Spingola2020b}
{Spingola}, C., {McKean}, J.~P., {Vegetti}, S., {et~al.} 2020, \mnras, 495,
  2387

\bibitem[{{Tavecchio} {et~al.}(2000){Tavecchio}, {Maraschi}, {Sambruna}, \&
  {Urry}}]{Tavecchio2000}
{Tavecchio}, F., {Maraschi}, L., {Sambruna}, R.~M., \& {Urry}, C.~M. 2000,
  \apjl, 544, L23

\bibitem[{{Urry} \& {Padovani}(1995)}]{Urry1995}
{Urry}, C.~M. \& {Padovani}, P. 1995, \pasp, 107, 803

\bibitem[{{Vito} {et~al.}(2019){Vito}, {Brandt}, {Bauer}, {Calura}, {Gilli},
  {Luo}, {Shemmer}, {Vignali}, {Zamorani}, {Brusa}, {Civano}, {Comastri}, \&
  {Nanni}}]{Vito2019}
{Vito}, F., {Brandt}, W.~N., {Bauer}, F.~E., {et~al.} 2019, \aap, 630, A118

\bibitem[{{Volonteri}(2012)}]{Volonteri2012}
{Volonteri}, M. 2012, Science, 337, 544

\bibitem[{{Volonteri} {et~al.}(2011){Volonteri}, {Haardt}, {Ghisellini}, \&
  {Della Ceca}}]{Volonteri2011}
{Volonteri}, M., {Haardt}, F., {Ghisellini}, G., \& {Della Ceca}, R. 2011,
  \mnras, 416, 216

\bibitem[{{Volonteri} {et~al.}(2015){Volonteri}, {Silk}, \&
  {Dubus}}]{Volonteri2015}
{Volonteri}, M., {Silk}, J., \& {Dubus}, G. 2015, \apj, 804, 148

\bibitem[{{Worrall} {et~al.}(2020){Worrall}, {Birkinshaw}, {Marshall},
  {Schwartz}, {Siemiginowska}, \& {Wardle}}]{Worrall2020}
{Worrall}, D.~M., {Birkinshaw}, M., {Marshall}, H.~L., {et~al.} 2020, \mnras
  [\eprint[arXiv]{2007.03536}]

\bibitem[{{Wright} {et~al.}(2010){Wright}, {Eisenhardt}, {Mainzer}, {Ressler},
  {Cutri}, {Jarrett}, {Kirkpatrick}, {Padgett}, {McMillan}, {Skrutskie},
  {Stanford}, {Cohen}, {Walker}, {Mather}, {Leisawitz}, {Gautier}, {McLean},
  {Benford}, {Lonsdale}, {Blain}, {Mendez}, {Irace}, {Duval}, {Liu}, {Royer},
  {Heinrichsen}, {Howard}, {Shannon}, {Kendall}, {Walsh}, {Larsen}, {Cardon},
  {Schick}, {Schwalm}, {Abid}, {Fabinsky}, {Naes}, \& {Tsai}}]{Wright2010}
{Wright}, E.~L., {Eisenhardt}, P. R.~M., {Mainzer}, A.~K., {et~al.} 2010, \aj,
  140, 1868

\bibitem[{{Wu} {et~al.}(2015){Wu}, {Wang}, {Fan}, {Yi}, {Zuo}, {Bian}, {Jiang},
  {McGreer}, {Wang}, {Yang}, {Yang}, {Thompson}, \& {Beletsky}}]{Wu2015}
{Wu}, X.-B., {Wang}, F., {Fan}, X., {et~al.} 2015, \nat, 518, 512

\bibitem[{{Zaroubi}(2013)}]{Zaroubi2013}
{Zaroubi}, S. 2013, Astrophysics and Space Science Library, Vol. 396, {The
  Epoch of Reionization}, ed. T.~{Wiklind}, B.~{Mobasher}, \& V.~{Bromm}, 45

\bibitem[{{Zhu}(2018)}]{Zhu2018}
{Zhu}, S. 2018, in American Astronomical Society Meeting Abstracts, Vol. 231,
  American Astronomical Society Meeting Abstracts \#231, 123.06

\end{thebibliography}

\appendix

\section{Derivation of physical parameters}
\label{sec:appendix_formulae}

We adopt the definition of synchrotron spectral index $\alpha$ as $S_{\nu} \propto \nu^{\alpha}$, where $S_{\nu}$  is the flux density at the observing frequency $\nu$. The uncertainty on the spectral index ($\sigma_{\alpha}$) between two frequencies $\nu_1$ and $\nu_2$ is given by 

\begin{equation}\label{eq:sigma_spix}
    \sigma_{\alpha} = \frac{1}{\ln(\nu_1/\nu_2)} \sqrt{\left(\frac{\sigma_{S_{\nu_1}}}{S_{\nu_1}}\right)^2 + \left(\frac{\sigma_{S_{\nu_2}}}{S_{\nu_2}}\right)^2 } ;
\end{equation}

where $\sigma_{S_{\nu_1}}$ and $\sigma_{S_{\nu_2}}$ are the uncertainties on the flux densities measured at the frequencies $\nu_1$ and $\nu_2$.

To estimate the possible ranges of viewing angles $\theta$ and of the bulk velocity in terms of speed of light $\beta_{\rm bulk}$ it is possible to use the jet/counter-jet ($J$) brightness ratio:

\begin{equation}\label{eq:jet_cj_viewing_angle}
    \beta_{\rm bulk}\cos(\theta) = (J^{1/p}-1)/(J^{1/p}+1) \; ;
\end{equation}

where $p = 2- \alpha$ assuming continuous and symmetric jets \citep[e.g.,][]{Giovannini1994} and $\alpha$ is the synchrotron spectral index of component 1a between 1.5 and 5 GHz (Fig.~\ref{fig:VLBI_spectrum}).
The Lorentz factor is related to the bulk velocity $\beta_{\rm app}$ as

\begin{equation}\label{eq:lorentz_factor}
   \Gamma = (1-\beta_{\rm bulk}^2)^{-1/2} \; ;
\end{equation}

and the Doppler factor $\delta$ is related to $\Gamma$ via

\begin{equation}\label{eq:doppler_factor}
    \delta = [\Gamma(1 -  \beta\cos\theta)]^{-1} .
\end{equation}

The source-frame brightness temperature is defined as

 \begin{equation}\label{eq:brightness_temperature}
    T_B(\lambda) = \frac{2 \ln 2}{\pi \kappa}  \frac{S_{\lambda} \lambda^2 (1+z)}{\theta_{\rm maj}\theta_{\rm min}}
    \; [{\rm K}]\; ;
\end{equation}

where $\kappa$ is the Boltzmann constant, $S_{\lambda}$ is the integrated flux density of each sub-component at the observing wavelength $\lambda$, $z$ is the redshift, $\theta_{\rm maj}$ and $\theta_{\rm min}$ are the FWHMs of the elliptical 2D Gaussian fits along the major and the minor axes, all in cgs units \citep[e.g.,][]{Kovalev2005}. 

By assuming that the contributions of the magnetic field  and the relativistic particles are equal \citep{Pacholczyk1970}, it is possible to estimate the so-called \textsl{equipartition magnetic field} as

\begin{equation}\label{eq:equipartition_magnetic_field}
H_{\rm eq} =  \left( 4.5 \; (1+\eta) \; c_{12} \; \frac{L}{V}\right)^{2/7} \; [{\rm G}] \; ;
\end{equation}

where we assumed $\eta =1$, meaning that proton and electron energies are assumed to be the same, $ c_{12}$ is a constant with value of $3.9\times 10^7$, $V$ is the volume of the radio-emitting source in cm$^3$ (that we assumed homogeneously filled by relativistic plasma), and $L$ is the bolometric radio luminosity (erg s$^{-1}$) between $\nu_1=10$~MHz and $\nu_2=100$~GHz defined as

\begin{equation}\label{eq:bolom_luminosity}
L = 4 \pi D^2_L \; (1+z)^{-(1+\alpha)}  \int_{\nu_1}^{\nu_2} S_0 \left( \frac{\nu}{\nu_0} \right)^{\alpha} \,d\nu \; ;
\end{equation}
where $D_L$ is the luminosity distance, $z$ is the redshift, $S_0$ is the flux density at the observing frequency $\nu_0$.

The ratio between the radio (due to synchrotron, $L_{\rm radio}$) and X-ray (due to inverse Compton, $L_{\rm X-rays}$) luminosities can also give an estimate of the magnetic field $H$ \citep{Schwartz2002}:

\begin{equation}\label{eq:synch_IC_magn_field}
    \frac{L_{\rm radio}}{L_{\rm X-ray}} = \frac{H^2/8\pi}{\rho_{\rm CMB}} \; 
\end{equation}

where $\rho_{\rm CMB} =  7.56\times10^{-15} T_0^4\; (1+z)^4 \;$ (erg cm$^{-3}$) is the energy density of the CMB, with $T_0 = 2.7255$~K is the value of the CMB temperature today \citep{Fixsen2009}.

The equivalent magnetic field at a redshift $z$ due to the CMB can be estimated following \citet{Ghisellini2014}:

\begin{equation}\label{eq:CMB_magn_field}
    B_{\rm CMB} = 3.26 \times 10^{-6} \; (1+z)^{2} \; [G].
\end{equation}

\section{VLA and VLBA observations properties}
\label{sec:appendix_tables}

In this section, we report the parameters of the self-calibrated VLBA images (Table~\ref{tab:images_properties}), the properties of the VLBI sub-components (Table~\ref{tab:vlbi_properties}) and the VLA measurements (Table~\ref{tab:appendix_vla_data}).

\begin{table*}[ht]
    \centering
   \caption{Properties of the self-calibrated VLBA images of PSO~J0309+27 shown in Fig.~\ref{fig:VLBA_images}, such as the observing frequency (column 1), total integrated flux density (column 2), Peak surface brightness (column 3), off-source rms noise (column 3), major and minor axes of the Gaussian restoring beam (column 4 and 5) and its position angle (column 6, east of north).} \label{tab:images_properties}
    \begin{tabular}{lllllll}
    \hline
        Freq. & Total flux density & Peak surface brightness & RMS & $B_{\rm maj}$ & $B_{\rm min}$ & $B_{\rm PA}$  \\
       (GHz) & (mJy) & (mJy beam$^{-1}$) & ($\mu$Jy beam$^{-1}$) & (mas) &  (mas) & (deg) \\
       \hline
        1.5 & $20\pm2$ & $7.2\pm0.7$ & 42 & 12.8 & 7.6 & 0  \\
        5.0 & $7.0\pm0.7$ & $4.1\pm0.4$ & 25  & 4.2 & 2.9 & 14\\
        8.4 & $5.1\pm0.5$ & $3.2\pm0.3$ & 43 & 2.8 & 2.3 & 23 \\
    \hline
    \end{tabular}
\end{table*}
\bigskip

\begin{table*}
 \centering 
  \caption{Spatially resolved radio properties of PSO~J0309+27 obtained using 2D Gaussian fitting: observing (and rest-frame in parentheses) frequency in GHz (column 1);  sub-component identification as labelled in Fig.~\ref{fig:VLBI_spectrum} (column 2);  peak surface brightness in mJy beam$^{-1}$ (column 3);  integrated flux density in mJy (column 4); elliptical size convolved by the beam, i.e. major and minor axes ($\theta_{\rm maj}$ and $\theta_{\rm min}$) in mas (column 5 and 6); rest-frame brightness temperature $T_B$ in units of $10^8$ K (column 7). If the subcomponent is unresolved, then we provide only lower limits on its $T_B$ (indicated by $>$ symbol). 
  }  \label{tab:vlbi_properties}

    \begin{tabular*}{0.9\textwidth}{lllllll}
\hline
 Frequency & Component  & Peak flux density  & Integrated flux density   & $\theta_{\rm maj}$ & $\theta_{\rm min}$  & $T_B$ \\
  (GHz)  &  & (mJy beam$^{-1}$)  & (mJy)& (mas) & (mas)  &  ($10^8$ K)   \\     
\hline
\multirow{4}{*}{1.5 (10.65)} & 1 (core) & $6.83 \pm 0.04$ & $8.02 \pm 0.08$ & $13.6\pm0.1$ &$8.3\pm0.1$  & $2.7\pm0.2$ \\
& 2  & $4.82 \pm 0.04$  & $5.80 \pm 0.08$ & $13.6\pm0.2$ & $8.6\pm0.1$ & $1.9\pm0.2$\\
& 3 & $1.98 \pm0.04$  & $4.03 \pm 0.12$  & $19.2\pm0.5$ & $10.3\pm0.5$ & $0.8\pm0.1$ \\
& 4 & $0.30 \pm 0.04 $ & $1.78 \pm 0.25$ & $20.5\pm0.5$ & $16.1\pm1.1$ & $0.2\pm0.1$ \\
\hline
\multirow{4}{*}{5 (35.5)} & 1a (core)  & $4.18 \pm 0.03$ & $4.29 \pm 0.05$ & $4.2\pm0.1$ & $2.9\pm0.1$ & $>1.19$ \\
& 1b & $0.70 \pm 0.03$  & $0.74 \pm 0.05$ & $4.2\pm0.2$ & $2.9\pm0.1$ & $>0.20$ \\
& 2  & $1.02 \pm 0.03$ &  $1.41 \pm 0.06$ & $5.4\pm0.2$ & $3.0\pm0.1$ & $0.30\pm0.02$ \\
& 3  & $0.12 \pm 0.03$ & $0.60 \pm 0.15$  & $12.3\pm3.1$ & $ 4.9\pm0.9$ & $0.03\pm0.01$ \\
\hline
\multirow{5}{*}{8.4 (59.64)} & 1a (core) & $3.21 \pm 0.04$ &  $3.55 \pm 0.08$ & $2.8\pm0.1$ & $2.3\pm0.1$ & $>0.67$  \\
& 1b  & $0.46 \pm 0.04$ & $0.69 \pm 0.10$ &  $4.4\pm0.6 $ & $2.3\pm0.2$ & $0.09\pm0.01$ \\
& 2a & $0.55 \pm 0.04$ & $0.71 \pm 0.09$  & $3.1\pm0.3$ & $2.6\pm0.2$ & $0.11\pm0.02$ \\
& 2b & $0.23 \pm 0.04$ & $0.14 \pm 0.05$ & $2.3\pm0.5$ & $2.0\pm0.3$ & $0.04\pm0.02$ \\
\hline
    \end{tabular*} \medskip
\end{table*}

\begin{table*}
 \centering 
  \caption{ Observing frequency (column 1), flux density (column 2) and nominal uncertainty on flux density from 2D Gaussian fit (column 3) of PSO~J0309+27 measured from the VLA observations (Sec.~\ref{sec:observations} and Fig.~\ref{fig:vla_spectrum}).
  }  \label{tab:appendix_vla_data}
\begin{tabular}{lll|lll}
$\nu$ & $S_{\nu}$ & $\sigma_{S_{\nu}}$ & $\nu$ & $S_{\nu}$ & $\sigma_{S_{\nu}}$  \\
(GHz) & (mJy) & (mJy) & (GHz) &  (mJy) &  (mJy) \\
\hline
1.040   &  41.61    &  	0.75    & 12.082 &  3.81 &  0.07   \\
1.104   &  41.25	&   1.47    & 12.203 &  3.66 & 0.07   \\
1.170   &  \textsl{flagged}  &           & 12.331 &  3.44 & 0.07   \\
1.237   &  29.65	&   1.50     & 12.459 &  3.43 & 0.07   \\
1.298	&  \textsl{flagged}  &           & 12.587 &  3.71 & 0.07   \\
1.362   &  32.26	&  0.63     & 12.716 &  3.46 & 0.07   \\
1.424   &  30.42	&  0.28     & 12.843 &  3.46 & 0.07   \\
1.487   &  	28.75	&  0.30	    & 12.971 &  3.39 & 0.07   \\
1.541   &  \textsl{flagged}	&           & 13.099 &  3.53 & 0.07   \\
1.604   &  \textsl{flagged}  &  	        & 13.227 &  3.37 & 0.07   \\
1.679	&  24.97    &   0.32    & 13.355 &  3.34 & 0.06   \\
1.744	&  23.38    &  	0.29    & 13.476 &  3.29 & 0.06   \\
1.807	&  22.56    &  	0.29    & 13.618 &  3.23 & 0.06   \\
1.872	&  19.01    &  	0.27    &13.739  &  3.51 & 0.06   \\
1.917   &  11.21    &  	0.15    & 13.867 &  3.33 & 0.06   \\
1.974   &  12.22    &  	0.49    & 13.995 &  3.20 & 0.06   \\
2.022   &  21.00    &  	0.21    & 14.123 &  3.22 & 0.06   \\
2.179   &  \textsl{flagged}  &  	        & 14.251 &  3.06 & 0.07   \\
2.307   &  \textsl{flagged}  &           & 14.379 &  3.20 & 0.06   \\
2.396   &  18.10	&   0.18    & 14.507 &  3.16 & 0.06   \\
2.527   &  17.82	&   0.18    & 14.635 &  2.83 & 0.06   \\
2.655   &  16.59	&   0.17    & 14.763 &  2.92 & 0.06   \\
2.783	&  15.81    &  	0.17    & 14.891 &  3.12 & 0.06   \\
2.911	&  15.39    &  	0.16    & 15.012 &  2.69 & 0.07 \\
3.022	&  14.76    &  	0.16    & 15.154 &  2.90 & 0.07  \\
3.143	&  14.05    &  	0.15    & 15.275 &  2.69 & 0.07   \\ 
3.271	&  13.66    &  	0.14    & 15.403 &  2.77 & 0.07   \\
3.399	&  13.38    &  	0.14	& 15.531 &  2.81 & 0.07  \\
3.527	&  12.64    &  	0.08	& 15.658 &  2.70  & 0.07  \\
3.655	&  11.94	&   0.09	& 15.787 &  2.69 & 0.07   \\
3.783   &  11.56	&   0.10	& 15.915 &  2.51 & 0.07   \\
3.947	& \textsl{flagged}  &  	        & 16.043 &  2.77 &0.08   \\
4.522	&  9.44     &  	0.13    & 16.171 &  2.57 &0.07   \\ 
4.643	&  9.70     &  	0.06    & 16.299 &  2.51 & 0.07   \\
4.771	&  9.58     &  	0.08	& 16.427 &  2.53 & 0.07   \\
4.899   &  8.99     &  	0.08	& 16.548 &  2.74 & 0.07   \\
5.027   &  9.04	    &   0.08	& 16.690 &  2.64 & 0.07   \\
5.155   &  8.82	    &   0.08	& 16.810 &  2.48 & 0.08   \\
5.283   &  8.46	    &   0.09	& 16.939 &  2.48 & 0.08  \\
5.411   &  8.28     &  	0.08	& 17.067 &  2.48 & 0.08   \\
5.522   &  7.94	    &   0.10	& 17.195 &  2.50 & 0.08   \\
5.643   &  7.97     &  	0.08	& 17.323 &  2.32  & 0.08  \\
5.773   & \textsl{flagged}  &           & 17.451 &  2.37 & 0.09  \\
5.899   &  7.99	    &   0.08	& 17.579 &  2.27 & 0.08  \\
6.026   &  7.62	    &   0.09	& 17.707 &  2.14 & 0.08  \\
6.154   &  \textsl{flagged}  &  	        & 17.834 &  2.16 & 0.09  \\
6.282   &  7.04     &  	 0.09	& 17.963 &  2.22 & 0.10  \\
6.413   &  7.47     &    0.09	& 18.084 &  2.17 & 0.10  \\
8.051   &  5.64     &  0.06     & 18.695 &  2.13 & 0.05  \\
8.179   &  5.80     &   0.06    & 19.703 &  2.06 & 0.05  \\
8.307   &  5.57     &   0.06    & 20.695 &  2.01 & 0.05  \\
8.435   &  5.44     &   0.06    & 21.703 &  1.69 & 0.06   \\
8.563   &  5.37     &   0.06    & 22.695 &  1.68 & 0.06  \\
8.691   &  5.18     &   0.06    & 23.703 &  1.58 & 0.06  \\
8.819   &  5.06     &   0.06    & 24.695 &  1.52 & 0.06  \\
8.947   &  5.03     &   0.07    & 25.701 &  1.19 & 0.06   \\
9.058   &  5.08     &   0.06    & 30.995 &  1.05 & 0.07  \\
9.179   &  4.93     &   0.06    & 32.003 &  0.90 & 0.08  \\
9.303   &  4.93     &  0.06     & 32.995 &  1.02 & 0.08  \\
9.435   &  4.88     &   0.06    & 34.003 &  0.96 & 0.08  \\
9.563   &  4.83     &   0.06    & 34.995 &  0.93 & 0.08  \\
9.692   &  4.81     &   0.06    & 36.003 &  0.77 & 0.10   \\
9.819   &  4.81     &   0.06    & 36.995 &  0.89 & 0.10  \\
9.940   &  4.40     &   0.07    & 38.003 &  0.69 & 0.13   \\

\hline

\end{tabular}
\end{table*}

\clearpage

\end{document}